\documentclass{article}
\usepackage{amsmath, amssymb, amsthm}
\usepackage{physics}      
\usepackage{url}          
\usepackage{natbib}       

\usepackage[text={7in,9.6in}]{geometry}

\usepackage{rotating}

\begin{document}

\title{Bridging Classical Sensitivity and Quantum Scrambling: A Tutorial on Out-of-Time-Ordered Correlators}
\author{Stephen Wiggins \\
  Hetao Institute of Mathematics and Interdisciplinary Sciences, \\
  Floors 6-8, Building F, Shenzhen-Hong Kong International Science \& Technology Park, \\
  No. 3 Binglang Road, Futian Free Trade Zone, \\
  Shenzhen, Guangdong Province, China \\
  \and
  Emeritus Professor, School of Mathematics, University of Bristol, \\
  Fry Building, Woodland Road, Bristol, BS8 1UG, United Kingdom \\
  \texttt{stephen.wiggins@me.com}
}
\date{}

\maketitle

\begin{abstract}
In classical dynamical systems, chaotic behavior is often associated with exponential sensitivity to initial conditions together with global phase-space structure. Translating this geometric concept to the strictly linear framework of quantum mechanics presents a conceptual puzzle. The out-of-time-ordered correlator (OTOC) is often motivated as the quantum analogue of the classical butterfly effect, but this slogan can hide important mathematical distinctions. This tutorial bridges the gap between applied mathematics and quantum information by detailing the mathematical machinery of the OTOC. We explore how classical sensitivity translates to operator non-commutativity, why standard two-point correlation functions fail to cleanly detect this sensitivity, and how the delocalization of quantum observables relates to classical notions of mixing. Crucially, we outline what the OTOC can and cannot diagnose, distinguishing between local instability and global chaos. Ultimately, we provide a precise and usable conceptual map, exploring how the Koopman-von Neumann formalism offers a framework to view classical and quantum dynamics through a shared linear perspective.
\end{abstract}


\section{Introduction}
In classical dynamical systems, chaotic behavior is associated with sensitive dependence on initial conditions together with global mechanisms that repeatedly stretch and fold phase-space structure. The exponential divergence of nearby trajectories---the ``butterfly effect''---is traditionally measured via Lyapunov exponents. However, transferring these geometric concepts to quantum mechanics presents an apparent tension: the Schr\"odinger equation is strictly linear, and unitary evolution preserves inner products, seemingly obstructing a direct translation of classical topological mixing into quantum mechanics.

In recent years, the physics and applied mathematics communities have converged on a mathematical diagnostic to address this: the Out-of-Time-Ordered Correlator (OTOC). In high-energy physics, OTOCs have been used to argue that black holes are the fastest scramblers of information in the universe, providing a structural link between quantum chaos and holographic string theory \citep{roberts2015diagnosing, maldacena2016bound}. 

In the laboratory, OTOCs have emerged as a valuable probe of operator spreading and information scrambling in quantum many-body systems \citep{dalessio2016from, lewis_swan2019dynamics}. By probing how an initially localized observable dynamically couples to many degrees of freedom, researchers can access complex quantum dynamics and distinguish them from simpler environmental noise and decoherence. This paper provides a careful introduction to OTOCs for an audience of applied mathematicians and dynamicists. We detail how classical sensitivity is algebraically translated into operator non-commutativity, why standard correlation functions fail to detect scrambling, and how to safely navigate the boundaries between local instability, global chaos, and quantum entanglement.

\section{Mathematical Preliminaries: Quantum Mechanics for Dynamicists}
To construct the OTOC, we must first translate classical phase space variables into the language of linear operators acting on a Hilbert space $\mathcal{H}$ \citep{dirac1981principles}.

\subsection{States and Observables}
In classical mechanics, a state is a point $(x,p)$ in phase space, and observables are real-valued functions $f(x,p)$. In quantum mechanics, a state is a complex vector in $\mathcal{H}$. Throughout this paper, we employ Dirac's standard ``bra-ket'' notation \citep{dirac1981principles}, which visually distinguishes a column vector (a ``ket'', denoted $\ket{\psi}$) from its dual row vector (a ``bra'', denoted $\bra{\psi}$). The inner product of two states is therefore written seamlessly as the bracket $\langle \phi | \psi \rangle$. Observables are elevated to \textbf{operators} acting on $\mathcal{H}$. To ensure measurement outcomes are strictly real, physical observables must be \textbf{Hermitian} (self-adjoint), meaning a given operator $W$ is equal to its own conjugate transpose: $W = W^\dagger$. (Note that while mathematicians typically use an asterisk $W^*$ to denote the adjoint matrix, physics convention universally employs the dagger $W^\dagger$) \citep{dirac1981principles}.

\subsection{Time Evolution and the Heisenberg Picture}
Time evolution in quantum mechanics is generated by the Hamiltonian operator $H$ (the total energy of the system). The solution to the Schr\"odinger equation yields a time-evolution operator $U(t) = e^{-iHt/\hbar}$. Because probability must be conserved, $U(t)$ is \textbf{unitary}, satisfying $U^\dagger U = I$. 

To understand how operator spreading works, we must look at the \textbf{Heisenberg picture} of quantum mechanics. In classical mechanics, it is natural to think of the state of the system (the phase space point) moving over time. In quantum mechanics, we can alternatively keep the state vector completely static at $t=0$ and force the \emph{operators} to absorb the time dependence. 

The time-evolved operator is defined as:
\begin{equation}
W(t) = U^\dagger(t) W(0) U(t) = e^{iHt/\hbar} W(0) e^{-iHt/\hbar}
\end{equation}
This is an exact operator identity. To a dynamicist reading the algebra from right to left, it is often helpful to describe $W(t)$ informally as a sequence: ``evolve forward in time, act with the perturbation $W(0)$, then evolve backward in time.'' However, this phrase should be strictly understood as a mathematical mnemonic for the algebraic formula above; it is not an actual Loschmidt echo protocol, and it does not imply that an experiment literally runs time backwards.

\section{The Classical Butterfly Effect and Dirac's Rule}
Classical sensitivity to initial conditions is defined by the partial derivative of a final position $x(t)$ with respect to an initial position $x(0)$. We can rewrite this derivative exactly in terms of the classical Poisson bracket \citep{haake2010quantum}:
\begin{equation}
\{x(t), p(0)\} = \frac{\partial x(t)}{\partial x(0)} \frac{\partial p(0)}{\partial p(0)} - \frac{\partial x(t)}{\partial p(0)} \frac{\partial p(0)}{\partial x(0)}
\end{equation}
Since initial position and momentum are independent variables, $\partial p(0) / \partial x(0) = 0$. Thus, the classical butterfly effect can be directly encoded in a Poisson bracket:
\begin{equation}
\frac{\partial x(t)}{\partial x(0)} = \{x(t), p(0)\}
\end{equation}

In a chaotic system, this derivative grows exponentially as $e^{\lambda_L t}$, where $\lambda_L$ is the Lyapunov exponent \citep{haake2010quantum}. To import this to quantum mechanics, we apply \textbf{Dirac's quantization rule}, which maps the classical Poisson bracket algebra directly to the quantum commutator algebra \citep{dirac1981principles}:
\begin{equation}
\{A, B\} \longrightarrow \frac{1}{i\hbar} [A, B]
\end{equation}
where the commutator is explicitly defined as $[A, B] = AB - BA$. This provides a leading-order correspondence in the semiclassical regime. The quantum analogue of classical sensitivity is therefore the commutator of time-separated operators: $[x(t), p(0)]$. For generic local operators $W$ and $V$, the footprint of instability is the algebraic growth of their commutator $[W(t), V(0)]$ \citep{larkin1969quasiclassical, swingle2018unscrambling}.

\section{The Necessity of the 4-Point OTOC}

To understand why physicists rely on a ``4-point'' correlator, we must first look at how statistical properties are typically analyzed in classical dynamical systems.

\subsection{The Failure of 2-Point Functions}
In classical statistical mechanics and signal processing, we study the dynamics of an observable $A$ by computing its autocorrelation function, $\langle A(t) A(0) \rangle$. (Note that while dynamicists often use angle brackets to denote time or phase-space averages, in this quantum context, the angle brackets denote the \textbf{quantum expectation value} of an operator. For a system in a perfectly isolated pure state $\ket{\psi}$, this expectation value for the observable $A$ is precisely the inner product $\langle \psi | A | \psi \rangle$. For a macroscopic system at temperature $T$, it represents a statistical average weighted by the quantum Gibbs measure $\rho \propto e^{-H/k_B T}$, computed via the matrix trace $\text{Tr}(\rho A)$.) 

Because this formula measures the relationship between the system at exactly two distinct points in time, physicists refer to it as a \textbf{2-point correlation function}. A 2-point function is highly useful for measuring \emph{relaxation} or \emph{dissipation}---how quickly a system forgets its initial state and settles into thermal equilibrium. However, it cannot measure the \emph{butterfly effect}. Relaxation and scrambling are fundamentally different physical phenomena. A system can relax quickly without being chaotic (e.g., a heavily damped harmonic oscillator), and a system can be highly chaotic but take a long time to globally relax.

If we attempt to measure quantum sensitivity by simply computing the standard expectation value (the first moment) of the commutator, $\langle [W(t), V(0)] \rangle$, we run into a problem. Because it is not positive definite, it can vanish for symmetry reasons or suffer from severe phase cancellations even when the underlying operator spreading is extensive. Therefore, the first moment does not provide a robust measure of the magnitude of operator growth.

\subsection{Squaring the Commutator: The Birth of the 4-Point Function}
To extract a robust, norm-like quantity that measures the absolute magnitude of this non-commutativity, we first define the positive-definite quantity:
\begin{equation}
C(t) = \frac{1}{\hbar^2} \langle [W(t), V(0)]^\dagger [W(t), V(0)] \rangle
\end{equation}
For Hermitian operators $W$ and $V$, where $W^\dagger = W$ and $V^\dagger = V$, this reduces to the more common form:
\begin{equation}
C(t) = -\frac{1}{\hbar^2} \langle [W(t), V(0)]^2 \rangle
\end{equation}

Expanding this Hermitian form yields:
\begin{equation}
C(t) = -\frac{1}{\hbar^2} \langle (W(t)V(0) - V(0)W(t))(W(t)V(0) - V(0)W(t)) \rangle
\end{equation}

When we expand this product, the non-commutativity of the operators generates four distinct terms. Among the four terms, the out-of-time-ordered cross-terms encode the failure of the two operator orderings to agree. One of these cross-terms takes the form:
\begin{equation}
F(t) = \langle W(t) V(0) W(t) V(0) \rangle
\end{equation}
which is a standard OTOC convention in the Hermitian case. Let us leverage the Heisenberg picture we established in Section 2.2 to interpret this expectation value. If we explicitly expand the Heisenberg operators, we see that $F(t)$ operates through an alternating sequence of forward and backward time evolutions punctuated by operator insertions:
\begin{equation}
F(t) = \langle \psi | \underbrace{U^\dagger(t) W(0) U(t)}_{W(t)} V(0) \underbrace{U^\dagger(t) W(0) U(t)}_{W(t)} V(0) | \psi \rangle
\end{equation}

Recall that quantum operators act on the state vector $\ket{\psi}$ from right to left. We can conceptually split this sequence into the inner product (the overlap) of two operator-insertion sequences, $F(t) = \langle \phi_2 | \phi_1 \rangle$:
\begin{itemize}
    \item \textbf{Sequence 1 ($\ket{\phi_1} = W(t) V(0) \ket{\psi}$):} The operator $V(0)$ acts on the system at $t=0$. This can be read algebraically as if the state were then evolved \emph{forward} to time $t$, acted on by $W(0)$, and finally pulled \emph{backward} to the present by the adjoint evolution.
    \item \textbf{Sequence 2 ($\ket{\phi_2} = V(0) W(t) \ket{\psi}$):} This sequence can be read as if the state were evolved \emph{forward} to time $t$ entirely unperturbed, acted on by $W(0)$, pulled \emph{backward} to $t=0$, and \emph{finally} acted on by $V(0)$.
\end{itemize}

Because this inner product requires evaluating exactly four distinct operator insertions across these folded time paths, it is formally classified as a \textbf{4-point Out-of-Time-Ordered Correlator}. If the dynamics generate rapid operator growth, then the two orderings $W(t)V(0)$ and $V(0)W(t)$ produce increasingly different states. The decay of their overlap $F(t)$ is one way to interpret the growth of the squared commutator, providing a mathematical translation of noncommuting operator sequences into the linear algebra of quantum mechanics.

\section{What OTOCs Measure---and What They Do Not}
The OTOC is an incredibly powerful diagnostic, but it is frequently subjected to semantic drift in the literature. To build a reliable conceptual map, we must clearly distinguish between several related, but distinct, phenomena.

\subsection{The Mechanism of Operator Growth}
In classical mechanics, a multidimensional phase space is constructed via the Cartesian product of individual subsystems. In quantum mechanics, a many-body Hilbert space is constructed via the tensor product ($\mathcal{H}_1 \otimes \mathcal{H}_2 \otimes \dots$). Therefore, an initial perturbation $W(0)$ that is physically localized to a single subsystem is written mathematically as an operator acting on just that subspace, while acting as the identity $I$ (doing nothing) everywhere else:
\begin{equation}
W(0) = w_1 \otimes I \otimes I \otimes \dots \otimes I
\end{equation}

To see exactly how this local operator spreads, we expand the Heisenberg time evolution $W(t) = e^{iHt/\hbar} W(0) e^{-iHt/\hbar}$ using the Hadamard lemma (frequently referred to in physics literature as the Baker-Campbell-Hausdorff expansion) \citep{hall2015lie}:
\begin{equation}
W(t) = W(0) + \frac{it}{\hbar}[H, W(0)] - \frac{t^2}{2!\hbar^2}[H, [H, W(0)]] + \dots
\end{equation}

This infinite series of nested commutators is the literal mathematical engine of operator growth. Suppose the system is a one-dimensional chain of interacting particles, and the Hamiltonian $H$ contains nearest-neighbor coupling terms (e.g., interactions linking particle 1 to 2, 2 to 3, etc.).
\begin{itemize}
    \item The zeroth-order term, $W(0)$, acts only on site 1.
    \item In the first-order term, the commutator $[H, W(0)]$ invokes the coupling between site 1 and site 2, generating a new operator that acts non-trivially on both sites.
    \item In the second-order term, the nested commutator $[H, [H, W(0)]]$ couples those first two sites to site 3, generating an operator with support across three sites.
\end{itemize}

As time evolves, the higher-order terms become significant. The initially simple, sparse operator $W(0)$ dynamically morphs into a massive linear combination of dense tensor products acting on an increasingly large number of degrees of freedom.

\subsection{Growth vs. Entanglement vs. Scrambling}
While the OTOC successfully detects this operator growth, one should carefully separate the following concepts to avoid semantic drift:

\begin{itemize}
    \item \textbf{Operator Growth:} The algebraic expansion of an observable's spatial support via nested commutators, as derived above.
    \item \textbf{Scrambling:} The delocalization of this initially localized information across many degrees of freedom in the system.
    \item \textbf{Thermalization:} The process by which local observables relax to steady-state values that match statistical ensembles. Because unitary evolution and quantum Poincar\'e recurrence prevent a strict mathematical limit at $t \to \infty$,\footnote{In a closed quantum system with a discrete spectrum, unitary time evolution is distance-preserving and possesses no attractors. Therefore, the expectation value $\langle A(t) \rangle$ is strictly a quasi-periodic function. By the quantum Poincar\'e recurrence theorem \citep{bocchieri1957quantum}, the constituent phases will eventually realign, returning the system arbitrarily close to its initial state. Consequently, a strict asymptote cannot exist; thermalization instead implies the expectation value remains near equilibrium for the exceptionally long time prior to recurrence.} this relaxation does not imply a permanent asymptote; rather, it is rigorously understood to mean that the system remains arbitrarily close to the equilibrium value for almost all times \citep{dalessio2016from}.
    \item \textbf{Entanglement Entropy:} In classical mechanics, each subsystem of a composite system always possesses a well-defined individual state. In quantum mechanics, the tensor product structure allows for entangled states that cannot be factored into independent local states. To quantify this, a system is spatially partitioned into regions $A$ and $B$. By mathematically averaging over (tracing out) the degrees of freedom in $B$, one constructs the \textbf{reduced density matrix} $\rho_A = \text{Tr}_B(\ket{\psi}\bra{\psi})$. To a dynamicist, $\rho_A$ is the quantum analogue of a marginal probability distribution, capturing the statistical uncertainty in $A$ generated entirely by its entanglement with $B$. The \textbf{von Neumann entropy} is then the Shannon entropy of this matrix's eigenvalues: $S_A = -\text{Tr}(\rho_A \ln \rho_A)$ \citep{nielsen2010quantum}. This scalar measures exactly how much quantum information is non-locally shared across the partition boundary.
    \item \textbf{Entanglement Growth vs. OTOCs:} As a strongly interacting system evolves, this entanglement entropy grows dramatically. However, it is crucial to separate this state-based entropy from operator-based OTOCs. Mathematically, von Neumann entropy is a property of a state vector (or density matrix), whereas the squared commutator is an algebraic property of the observables themselves. Operator spreading (measured by the OTOC) is closely related to the generation of this entanglement \citep{lewis_swan2019dynamics}. While they exhibit correlated growth regimes in specific chaotic models, they represent fundamentally different mathematical objects and should not be conflated.
\end{itemize}

\subsection{Local Instability is Not Global Chaos}
A second critical distinction must be drawn between local instability and global chaos. In classical mechanics, trajectories near an unstable fixed point or a saddle will separate exponentially, even if the system as a whole is entirely integrable and possesses no global chaotic mixing. This exact same phenomenon occurs with the OTOC. Exponential OTOC growth can be generated purely by a local unstable region rather than fully developed global chaos. This has been clearly demonstrated in one-dimensional models like the inverted harmonic oscillator, where the unstable maximum drives a clean exponential OTOC window despite the system not being chaotic in the standard topological sense \citep{hashimoto2020exponential}. Therefore, the slogan ``OTOC growth equals chaos'' is an oversimplification; the more accurate statement is that OTOCs successfully detect instability and operator growth, but whether that instability reflects global chaos depends entirely on the macroscopic bounds of the model.

\section{Scrambling, the Ehrenfest Time, and Geometry}
To understand how the strictly linear framework of quantum mechanics relates to classical non-linear folding, it is illuminating to consider two classical analogies where linear motion becomes chaotic purely due to topological boundary conditions:

\begin{enumerate}
    \item \textbf{The Arnold Cat Map:} This map is a chaotic system driven by a purely linear transformation (a $2 \times 2$ matrix) \citep{arnold1968ergodic}. Unbounded linear stretching generates intense global chaos strictly because the map is defined on a torus, forcing the fluid to fold back upon itself.
    \item \textbf{Chaotic Billiards:} For a particle moving in a Bunimovich stadium \citep{bunimovich1979ergodic}, the internal motion is strictly linear. The chaotic mixing is generated entirely by the boundaries (the hard walls) reflecting the diverging trajectories.
\end{enumerate}

Quantum scrambling operates on a similar structural principle. The Schr\"odinger equation is strictly linear, while Heisenberg operators become increasingly complex under nested commutators. However, the system's geometric bounds break the scrambling process into distinct phases. It is here that we must carefully distinguish between finite and infinite dimensional vector spaces.

\begin{itemize}
    \item \textbf{The Stretching (Early Time):} The Heisenberg-evolved operator becomes increasingly complex under nested commutators, and in suitable regimes the corresponding OTOC may exhibit early-time exponential growth. If the classical limit of the system possesses a non-uniformly hyperbolic chaotic sea \citep{lichtenberg1992regular}, this geometry drives the early exponential growth.
    \item \textbf{The Macroscopic Boundary (The Ehrenfest Time):} The timescale at which this early-time operator growth breaks correspondence with classical trajectories is known as the \textbf{Ehrenfest time}, $t_E$. Its classical analog is the mixing time: the time it takes for a microscopic phase space fluid volume to stretch across the macroscopic domain. Because quantum mechanics bounds this minimum volume by the Heisenberg uncertainty principle, the Ehrenfest time often scales logarithmically with $1/\hbar$ \citep{berman1978condition, haake2010quantum}.
    \item \textbf{Finite-Dimensional Saturation (Hitting the Walls):} For spin chains and other strictly finite-dimensional many-body systems, saturation is inevitable. In finite-dimensional operator spaces, growth cannot continue indefinitely into new independent directions, so the OTOC eventually crosses over from early-time growth to saturation \citep{maldacena2016bound, luitz2017information}.
    \item \textbf{The Infinite-Dimensional Caveat:} Crucially, having a finite number of physical degrees of freedom does \emph{not} imply a finite-dimensional Hilbert space. Continuous-variable systems, such as harmonic oscillators or bosonic modes, possess infinite-dimensional Hilbert spaces. For these systems, the ``hitting the walls'' metaphor fails. The mechanism of late-time saturation, recurrence, or oscillation in continuous spaces is highly model-dependent and cannot be explained solely by finite-dimensional boundary geometries \citep{rozenbaum2017lyapunov, hashimoto2020exponential}.
\end{itemize}

\section{Bounding the Butterfly Effect: The MSS Bound}
While the mathematical structure of the OTOC bridges classical and quantum mechanics, the physical limits of the butterfly effect diverge drastically between the two regimes. Classically, there is no fundamental maximum speed limit on chaos; if a potential is made infinitely steep, the trajectories can diverge infinitely fast. In the quantum realm, Maldacena, Shenker, and Stanford established a fundamental speed limit on this growth, known as the MSS bound \citep{maldacena2016bound}. To rigorously define a quantum Lyapunov exponent $\lambda_L$ from an OTOC, a system must possess a vast separation of timescales. Specifically, there must be a wide intermediate window of time that is much longer than the local thermal dissipation time ($t_d \sim \hbar/k_B T$), yet much shorter than the macroscopic Ehrenfest time ($t_E$) where the correlator inevitably saturates. If a thermal quantum system possesses a sufficient number of interacting degrees of freedom (often denoted in physics as large-$N$ systems) to mathematically open this exponential window, the growth rate is strictly bounded by the temperature $T$:
\begin{equation}
\lambda_L \le \frac{2\pi k_B T}{\hbar}
\end{equation}

This inequality is of profound interest in theoretical physics because holographic large-$N$ models---such as black holes \citep{shenker2014black} or the Sachdev-Ye-Kitaev model \citep{kitaev2015simple, maldacena2016remarks}---natively saturate it, leading to their classification as the ``fastest scramblers'' in nature \citep{maldacena2016bound}. However, for an applied mathematician, this bound must be applied with extreme caution. It relies on specific analyticity and factorization hypotheses for thermal correlators. It is not a universal blanket theorem covering every operator, every state, and every finite quantum model. Generic small-scale laboratory systems often lack the timescale separation required to even unambiguously define $\lambda_L$, let alone saturate the bound. Furthermore, the MSS theorem should not be used as a blunt instrument to analyze one-dimensional saddle models or inverted oscillators, where the observed growth is tied to locally unstable regions rather than true many-body thermal scrambling.

\section{Resolving the Paradox: The Koopman-von Neumann Perspective}
We opened this tutorial with an apparent tension: how can the strictly linear Schr\"odinger equation capture the highly non-linear topological mixing of classical chaos? Modern applied mathematics possesses a formal framework that addresses this apparent contradiction: the Koopman-von Neumann (KvN) formulation \citep{koopman1931hamiltonian, vonneumann1932operatorenmethode, mauro2002topics}.

KvN mechanics rewrites classical non-linear dynamics directly in the language of Hilbert spaces. Classical probability amplitudes evolve under a Liouville operator, and the resulting evolution is entirely linear at the level of observables or densities. 

This framework is conceptually valuable because it explicitly demonstrates that linear operator evolution is not, by itself, incompatible with classical chaos. However, the KvN formulation should be viewed as a powerful \emph{perspective}, not as a proof that classical chaos and quantum scrambling are the exact same phenomenon. It does not collapse the distinction between classical mixing and quantum scrambling into a single theorem. Instead, by placing both systems on identical linear footing, the real differences between classical and quantum dynamics are isolated elsewhere: in the non-commutativity of the algebra (Dirac's rule), in the discrete versus continuous nature of the spectral structure, and in the fundamental limits of measurement and interference.

\section{Conclusion}
The OTOC provides a useful framework for relating geometric ideas from dynamical systems to operator growth in quantum mechanics. Its primary strength is translating a classical-looking question about sensitivity into a genuinely quantum question about the growth of noncommutativity. However, the geometric perspective offers an illuminating alternative lens. Currently, OTOCs are often used primarily as macroscopic diagnostics to observe how fast a system scrambles a perturbation. But a dynamicist naturally recognizes that a generic Hamiltonian phase space is rarely a featureless, uniformly mixing sea. Instead, it is a deeply inhomogeneous topological landscape---a highly structured architecture of regular KAM islands woven together by the intricate heteroclinic tangles of stable and unstable manifolds. If quantum scrambling is influenced by this underlying classical skeleton, it suggests possible routes for more controlled, localized manipulation. The groundwork for this shift has recently been laid by researchers explicitly merging these vocabularies:

\begin{itemize}
    \item \textbf{Quantum Shielding (Suppressing Chaos via KAM Islands):} Rozenbaum, Ganeshan, and Galitski \citep{rozenbaum2017lyapunov} demonstrated this by studying the quantum kicked rotor, establishing that OTOC growth is a local property. By initializing a quantum wavepacket entirely within a regular classical KAM island, they showed that the OTOC fails to grow exponentially, effectively shielding the quantum state from the surrounding chaotic sea.
    \item \textbf{Accelerating Entanglement (Riding the Unstable Manifolds):} Conversely, Xu, Scaffidi, and Cao \citep{xu2020does} showed that global chaos is not required for rapid scrambling; placing a quantum state exactly at a classical unstable fixed point is sufficient to drive exponential OTOC growth. Building on this, Mu\~noz-Arias, Deutsch, and Poggi \citep{munoz2023phase} demonstrated that the exponentially optimal path to prepare highly entangled states for quantum metrology is to actively drive the system along the separatrices connecting phase space saddles.
\end{itemize}

\subsection*{Outlook}
Looking forward, this dual-vocabulary approach suggests several interdisciplinary frontiers. Future research might explore how heteroclinic tangles in many-body Hamiltonians relate to the spatial routing of operator spreading across a quantum lattice. Ultimately, translating the topological language of classical chaos into the Hilbert space of quantum mechanics refines our understanding of the quantum butterfly effect. Ideally, a tutorial should leave the reader with both enthusiasm for these connections, and the mathematical caution required to navigate their limits.

\subsection*{Acknowledgements}
I am grateful to Arseni Goussev for comments on an earlier version of this manuscript.

\end{document}